\documentstyle[12pt]{article}
\title{A new class of non-Hermitian Hamiltonians with real spectra}
\author{T. V. Fityo}

\begin{document}
\maketitle
{\small
{\center
Ivan Franko National University of Lviv, Chair of Theoretical Physics \\
12 Drahomanov Str., Lviv UA-79005, Ukraine\\
fityo@ktf.franko.lviv.ua\\}}

{\abstract We construct a new class of non-Hermitian
Hamiltonians with real spectra. The Hamiltonians possess one
explicitly known eigenfunction.}

\section{Introduction}

By now, non-Hermitian Hamiltonians attract a lot of attention. Such
Hamiltonians are used
in optics \cite{swfi, dbvg}, in field theory \cite{ncos} and other branches
of theoretical physics.

Among the non-Hermitian Hamiltonians much attention was devoted to
investigation of properties of the so-called $PT$ symmetric
Hamiltonians \cite{eoch, rsin, ptsq, tcp, sswh, ssfp, ncop, copn,
gce,fspo}. A Hamiltonian is called to be $PT$ symmetric if
$PTH=HPT$, where $P$ is the parity operator, i.e. $Pf(x)=f(-x)$,
and $T$ is the complex conjugation operator. The main reason for
this interest was an assumption that their spectra were entirely
real as long as the $PT$ symmetry was not spontaneously broken.

There are several ways to build a non-Hermitian Hamiltonian with real
spectrum. For this purpose
it was proposed to use supersymmetric quantum mechanics \cite{sowc}.

Recently, Mostafazadeh generalized $PT$ symmetry by
pseudo-Hermiticity \cite{phvp}. The idea of pseudo-Hermiticity was
introduced by Dirac \cite{Dirac}. A Hamiltonian $H$ is said to be
$\eta$-pseudo-Hermitian if
\begin{equation}\label{ph}
H^+=\eta H\eta^{-1},
\end{equation}
where $^+$ denotes the operation of adjoint. In \cite{phvp} it was
proposed a new class of non-Hermitian Hamiltonians with real spectra
which are obtained using pseudo-supersymmetry.

Mostafazadeh also showed \cite{phvp2} that the operator $H$ with
complete set of biorthonormal eigenvectors has a real spectrum if
and only if there exists a linear invertible operator $O$ such that
$H$ is $\eta$-pseudo-Hermitian, where $\eta=O^+O$.

In the paper we construct a new class of pseudo-Hermitian operators with real
spectra using $O$ as a first order differential operator.

\section{Pseudo-Hermiticity}

Suppose that non-Hermitian Hamiltonian $H$ is $\eta$-pseudo-Hermitian:
\begin{equation}\label{e1}
\eta H =H^+\eta.
\end{equation}
Here, we choose another form of pseudo-Hermiticity to avoid a necessity of $\eta$
invertibility (the form (\ref{e1}) is mentioned in \cite{phvp}).

Choose an operator $\eta$ to be an Hermitian operator. Then $\eta H$ is an
Hermitian operator, too: $(\eta H)^+=H^+\eta^+=H^+\eta=\eta H$. Consider an
eigenfunction $\psi$ and the corresponding eigenvalue $E$ of $H$. Then, because
of Hermiticity of $\eta H$ as well as of $\eta$,
\begin{equation}
\int\psi^*\eta H\psi dx=E\int\psi^*\eta\psi dx,
\end{equation}
both integrals are real and except for the case
\begin{equation}\label{eq0}
\int\psi^*\eta\psi dx=0
\end{equation}
the eigenvalue $E$ is also real.
On the contrary, if $\int \psi^*\eta\psi dx=0$ then the left integral of
(\ref{eq0}) has to be zero, too. In this case $E$ can be either a real or
a complex number.

For a general form of $\eta$ it is difficult to find if there
exist such eigenfunctions which satisfy (\ref{eq0}). To simplify
the study of the case of $\int \psi^*\eta\psi dx=0$ we
concretize the form of $\eta$ to be
\begin{equation}\label{e3}
\eta=O^+O.
\end{equation}

For this case the integral $\int\psi^*O^+O\psi dx = \int |O\psi|^2dx$ is greater
than zero except for the case of $\psi$ belonging to the kernel of $O$. So we have
to solve
\begin{equation}\label{e4}
O\phi=0
\end{equation}
and verify if solutions of this equation are the eigenfunctions of $H$.

In the following section we build such a pair of Shr\"odinger Hamiltonian
\begin{equation}\label{ShH}
H=-\frac{d^2}{dx^2}+V(x)
\end{equation}
and $O^+O$ that satisfies condition (\ref{e1}).

\section{$O$ as the first order differential operator}\label{anc}

Choose $O$ in the following form
\begin{equation}\label{e5}
O=\frac{d}{dx}+f(x)+ig(x),
\end{equation}
where $f$, $g$ are regular, real-valued functions. Then
\begin{equation}\label{e6}
O^+=-\frac{d}{dx}+f(x)-ig(x).
\end{equation}

Substituting (\ref{ShH}), (\ref{e5}) and (\ref{e6}) into (\ref{e1}) and
collecting terms with $\frac{d^2}{dx^2}$ operator we obtain
\begin{equation}\label{e7}
{\rm Im} V=-2g'.
\end{equation}

The terms without differential operators lead to $4g'(f'+f^2)+2g(f'+f^2)'=g'''$.
Multiplying this equation by $g$ and integrating it we obtain
\begin{equation}\label{e9}
f^2-f'=\frac{2gg''-g'^2+\alpha}{4g^2},
\end{equation}
where $\alpha$ is a real constant of integration.

The terms with $\frac{d}{dx}$ give $2{\rm Re} V'=2(f^2-f'-g^2)'$.
Integrating it and substituting (\ref{e9}) we can rewrite the real part
of potential as
\begin{equation}\label{e8}
{\rm Re}V=f^2-f'-g^2+\beta=\frac{2gg''-g'^2+\alpha}{4g^2}-g^2+\beta,
\end{equation}
where $\beta$ is a real constant of integration. In equations
(\ref{e7}-\ref{e8}) $g$ plays a role of generating function. In
order to obtain a $PT$ symmetric Hamiltonian the generating
function $g$ must be an even function, i.e. $g(x)=g(-x)$.

It should be noted that the choice (\ref{e5}) leads to
$\eta=-\frac{d^2}{dx^2}-2ig\frac{d}{dx}+f^2-f'+g^2-ig'$ and $\eta$ plays a
role of a second order Darboux operator. It intertwines $H$ and $H^+$
which are superpartners of the second order supersymmetry \cite{dtfq}. Formulae
(\ref{e7}-\ref{e8}) are similar to the corresponding results of \cite{sssq}.

The next step is to check whether the solution of (\ref{e4}) is an eigenfunction
of $H$. In terms of $f$ and $g$ we can express this solution as:
\begin{equation}\label{e10}
\phi=e^{-\int (f+ig) dx}.
\end{equation}
Considering $\phi$ as an eigenfunction of (\ref{ShH}) and using
(\ref{e7}), (\ref{e8}) we obtain
\[-i(g'+2fg)+\beta=E,\]
where $E=E_r+iE_i$ is the complex eigenvalue of $H$ ($H\phi=E\phi$). We see
that $\beta=E_r$. Then
\begin{equation}\label{e111}
f=-\frac{E_i+g'}{2g}.
\end{equation}

Now,
from (\ref{e111}) and (\ref{e9}) we have two different relations between
$f$ and $g$. To compare them we substitute $f$ from (\ref{e111}) into
(\ref{e9}) and after some simplification we obtain $E_i^2=\alpha$. So we can state that
$\phi$ can be an eigenfunction of (\ref{ShH}) only if $\alpha\ge0$.
Note that (\ref{e111}) for the case $E_i^2=\alpha$ is the solution of
(\ref{e9}).

So choosing any $g$ and $\alpha<0$ we can be sure that the spectrum of the
corresponding Hamiltonian is entirely real, but we are not
sure that it is not empty. By choosing for $\alpha=0$ a suitable $g$ one can
construct the Hamiltonian with real spectrum and that also possesses one explicitly known
eigenfunction. Choosing $g$ and $\alpha>0$ we have to check if the corresponding
$\phi$ does not belong to $L_2$ space to obtain an Hamiltonian with real
spectrum.

In the following section we illustrate these results.

\section{Examples}

For constructing Hamiltonians we use formulae (\ref{e7}) and
(\ref{e8}) to represent imaginary and real part of the potential
as well as (\ref{e111}) to express $f$. There are two ways to obtain
regular expression for $f$. The first is to choose $g$ without
sign changing and any value of $E_i$ or $\alpha$. It is
illustrated with example 1. The second way is to choose $g$ as a
function with a simple zero. In this case we have to fix value of
the $E_i$ to avoid singularity. This way is illustrated with
examples 2 and 3.

{\em Example 1}

Choosing the generating function $g$ as the even function
\[g=e^{-x^2}\]
we obtain $PT$ symmetric Hamiltonian
\begin{equation}\label{eeeH}
H=-\frac{d^2}{dx^2}+x^2+\frac{\alpha}{4}e^{2x^2}-e^{-2x^2}-4ixe^{-x^2}+\beta-1
\end{equation}
which possesses real spectrum for $\alpha<0$, for $\alpha=0$ we know one
eigenfunction $\psi_{E=\beta}=\exp(-\frac{x^2}{2}-i\int e^{-x^2}dx)$ and for
$\alpha=E_i^2>0$ the eigenfunction
$\psi_{E=\beta+iE_i}=\exp(-\frac{x^2}{2}+\frac{E_i}{2}
\int e^{x^2}dx-i\int e^{-x^2}dx)$
does not belong to $L_2$ space. So we can
state that spectrum of (\ref{eeeH}) is entirely real for any value of
$\alpha$ parameter.

{\em Example 2}

Choose the generating function $g$ in the form
\[g=\sinh(x),\]
then, to obtain regular $f=-\frac{E_i+\cosh(x)}{2\sinh(x)}$ one must set
$E_i=-1$ and then $f=-\frac{1}{2}\tanh{\frac{1}{2}x}$. Then
$\phi=\cosh(\frac{1}{2}x)e^{-i\cosh(x)}$ does not belong to $L_2$. So
spectrum of
\[H=-\frac{d^2}{dx^2}-2i\cosh(x)-\sinh^2(x)\]
is real.

{\em Example 3}

Choose the generating function $g$ in the form
\[g=\tanh(x),\]
then, avoiding singularity, we set $E_i=-1$ and obtain
\begin{equation}\label{eqH3}
H=-\frac{d^2}{dx^2}-\frac{2i-\frac{1}{4}}{\cosh^2(x)}+\beta-\frac{3}{4}
\end{equation}
with eigenfunction
$\psi_{E=\beta-i}=\frac{1}{\sqrt{\cosh(x)}}e^{-i\ln(\cosh(x))}$.
The spectrum of (\ref{eqH3}) can be found using supersymmetric methods and
it easy to show that this eigenvalue is unique.

\section{Acknowledgement}
I am very grateful to V. M. Tkachuk for numerous  useful 
discussions.

\end{document}